\documentclass[a4paper,11pt]{article}
\pdfoutput=1 
\usepackage{jcappub} 
\usepackage[T1]{fontenc}

\usepackage{graphics}
\usepackage{rotating}
\usepackage{pstricks}
\usepackage{color}
\usepackage{amsfonts}
\usepackage{mathrsfs}
\usepackage{epsfig}
\usepackage{amsmath,amssymb,amsthm,graphicx,latexsym,braket}
\usepackage{ulem}

\newcommand{\be}{\begin{equation}}
\newcommand{\ee}{\end{equation}}
\newcommand{\bea}{\begin{eqnarray}}
\newcommand{\eea}{\end{eqnarray}}
\newcommand{\bml}{\begin{subequations}}
\newcommand{\eml}{\end{subequations}}
\newcommand{\bfig}{\begin{figure}}
\newcommand{\efig}{\end{figure}}

\newcommand{\del}{\delta}

\title{\textsc{\fontsize{25}{70}\selectfont \sffamily \bfseries Exploring Inflationary Initial
 State With Large Scale Structure Observations}}
\author[1]{Debabrata Chandra\note{Electronic address: {deb.iitdelhi@gmail.com}}}
\author[2]{{ and Supratik Pal}\note{Electronic address: {supratik@isical.ac.in}}}

\affiliation{\textit{Physics and Applied Mathematics Unit, Indian Statistical Institute, \\203 B.T. Road, Kolkata 700 108, India}}

\abstract{
We investigate for possible
constraints on inflationary initial state from Large Scale Structure (LSS)
observations. Using a model-independent framework, we 
build the template for the generic initial state and 
construct the matter power spectrum through time
evolution of the primordial power spectrum.
We then
make use of the LSS data separately from 
the Sloan Digital Sky Survey - Data Release 7 (SDSS-DR7) sample of
Luminous Red Galaxies (LRG) and WiggleZ Dark Energy Survey,
and explore the plausible
constraints on initial vacuum by applying Bayesian parameter estimation method
with Markov Chain Monte Carlo (MCMC) simulation.  Our analysis reveals that,
along with the usual Bunch-Davies vacuum,
non-Bunch-Davies initial states are also fairly allowed so far as present LSS data
is concerned.}

\begin{document}
\maketitle
\flushbottom

\section{Introduction}\label{intro}

With the relentless advancement in observational cosmology
in the last decade coupled with a few new cosmological missions chipping in,
we are entering an era of precision cosmology.
Observational cosmology is progressively imparting 
stringent constraints on the cosmological parameters
that in turn helps us understand the  physics of our universe better and better.
However, in spite of the profound success of observational cosmology,
we are still left with conclusive determination of couple of
fundamental parameters that are essential ingredients in
improving our understanding 
of the origin and dynamics of the universe.

Over the past few decades, inflationary paradigm, supported by observational data
from WMAP\cite{Wmap} and Planck\cite{Planck},  
explains the physics of the early universe quite satisfactorily. 
Among all its successes, inflationary paradigm provides an ingenious mechanism
to explain the formation of seeds of classical perturbations, that have imprints on the
cosmic microwave background (CMB) anisotropy and, subsequently in
Large Scale Structure (LSS) data \cite{Guth}.
However, in spite of all the successes of inflation, 
 there
are few pivotal questions that are haunting us for long.
Some of these yet-unanswered questions to the 
building blocks of inflationary
dynamics are the following: (i) What is the energy scale of
inflation? \cite{Mirbabayi} (ii) What is/are the behavior of the
(self) interactions  of the field(s) involved in driving
inflation? \cite{Bartolo} (iii) What is the exact nature of the
inflationary initial state? \cite{Martin,Chandra} This article is solely
devoted to the investigation for any possible answer to the last question 
from existing LSS surveys
by educing all the precious information about the initial state. 

Despite investing a plethora of quality research towards probing
the physics of inflation in both ways, theoretically and experimentally,
 the true behaviour of the initial condition of the
inflationary fluctuations is an enigma to the fraternity.
However, before all of these, one first needs to know whether or not
exploring the nature of the primordial initial state at all has
any significance. To be honest, the subject is itself debatable.
So, let us first point out two distinct perceptions on the
significance of probing inflationary initial state: (i) Because of the superluminal
exponential expansion of the universe during inflation, the signature
of any non-trivial initial condition should be washed away,
and all we are left with is the Bunch-Davies (BD) vacuum which is nothing
but the lowest energy state \cite{Davies}, and is hence trivial; (ii) Due to the
underlying physical
processes, once the quantum fluctuations choose to raise its vacuum
to some non-trivial initial state, then its impression doesnot get faded
away, 
if the primordial fluctuations were linear in nature,
thereby BD vacuum is not restored. Reason behind
this is quite obvious, the role of initial condition is very
straightforward in linear perturbation theory, it comes into play
through the solution of the differential equation of the inflaton
fluctuations,  aka Mukhanov-Sasaki equation, representing
the equation of motion of the inflaton fluctuations, which get
reflected directly on the CMB and LSS through the evolution of
the inflaton perturbations. 
Thus any sort of customization on the
inflaton initial condition would be directly imprinted on the
early universe and late time physics \cite{Easther} through
evolution of the primordial perturbations.
As it stands, these are two conflicting statements and the debate is yet to
be settled.

One should however  note that between the two different
 views on the importance of primeval
initial condition the non triviality of the initial condition
have drawn more attention of the community, because of several reasons:
 (i) Just adequate amount of inflation may cause
to deflect the initial state from the BD vacuum to a non-trivial vacuum,
because, if the duration of  inflation were just as enough as to solve
the cosmological horizon and homogeneity problem then the largest modes
we are seeing today should generate from non-trivial initial state,
before inflation got settled to its attractor \citep{Wang}.
(ii) Some features (eg. periodic features) in inflationary model
may cause deviation of inflationary initial state from BD to a
non-trivial one \cite{chen2}. (iii) A different background
dynamics (eg. fluid dominated era) before inflationary epoch
allows inflation to occur from a initial state other than BD
(with non zero particle) \cite{Gasperini}. (iv) Trans-Planckian
physics can also act as one of the reason for compelling inflation to
start from a non vacuum initial state \cite{danielsson}.
(v) Multifield inflationary models too posses non Bunch-Davies(NBD) initial
state \cite{shiu}. 
On top of that, from theoretical point of view both BD and NBD
initial state have sufficient motivations to come up with.
So, it is crucial to know more about the exact nature of initial condition
making use of any possible imprints on observational data.

Our primary goal is to do a systematic, stepwise, model-independent development 
that would possibly help us investigate
 the nature of the primordial initial condition
from observations,  without giving priority to any specific
theoretical mechanism to set individual 
initial states.
  In an earlier work, we have already
made some progress in this direction by using
the latest Cosmic Microwave Background
(CMB) observations from Planck 2015 data \cite{Chandra},
and have found the possible  existence of non-standard
initial state of the inflaton fluctuations along with the usual BD vacuum. 
In this current work  we would like to progress in the same vein
and make use of a different observation, namely, the Large Scale Structure (LSS)
observations, to complement
our knowledge elicited from CMB.

LSS has tremendous potential to tell us about the initial
condition and the successive evolution of the
universe \cite{Alvarez}. LSS data provide us with ample
amount of information about the structure formations.
However, extracting primordial
information from LSS data is a very grueling venture,
because, there are plenty sources of late-time non-linearities
and noise, which makes the job quite arduous. Yet there are few
advantages that makes it a very tempting avenue for
the community, such as LSS data contain much more
information in comparison to CMB as it probes far more
number of scales than CMB\cite{Carrasco}. Apart from that,
LSS observations bestow us the opportunity to get a glimpse
of the evolution of our universe by providing a redshift
dependent data set, which allows us to get a 3 dimensional (3D) view
of the scales through redshift binning of the observed sky instead
of getting a snapshot of a particular epoch like CMB sky, that
is the snapshot of the last scattering surface. Hence, we
intend to study the impression of primordial initial state
on LSS, thereby try to constrain the nature of primordial initial state from LSS.
To this end, we will make use of two different LSS dataset, namely,
(i) the Sloan Digital Sky Survey - Data Release 7 (SDSS-DR7) sample of
Luminous Red Galaxies (LRG)\cite{SDSS} and (ii) WiggleZ Dark Energy Survey\cite{WZ}.
We will perform two separate analysis by using these two datasets
separately, that will also help us in having a comparative analysis
of the results from two different LSS datasets.

The plan of the paper is as follows: In Section 2,
we will describe a generic, model-independent parameterization of
the primordial power spectrum that takes into account
both BD and any non-trivial initial
state. Section 3 is dedicated to give a very brief {\it textbook} review
on the building of the dark matter power spectrum template
from primordial power spectrum, so as to correlate with our analysis
that follows is the next Section along with the data used in our analysis.
In Section 5 we discuss the results from our analysis, separately for WiggleZ and SDSS-DR7 LRG data.  
Finally we summarize our results with possible future directions.

\section{Parameterization of the power spectrum}\label{sec2}
In this section we will discuss about the template of the
inflationary power spectrum, which we are using throughout
our analysis to constrain the initial state from LSS,
consistent with the notations of this article.
A generic parameterization for the primordial power spectrum
of scalar fluctuations can be given by

\be
P(k)_{NBD}=\frac{k^3}{2 \pi^2}\left |\zeta\left(k\right)\right|^{2}
=P(k)_{BD}\left[|\alpha_{k}|^2+|\beta_{k}|^2-2Re\left(\alpha_{k}
 \beta_{k}^* \right)\right]
\label{nbdpwr} \ee
where, $\zeta$ stands for comoving curvature perturbation,
 $P(k)_{NBD}$ and $P(k)_{BD}$ are the power spectrum for
  non Bunch-Davies and Bunch-Davies initial condition respectively.
The power spectrum for Bunch-Davies
initial condition is defined as 
\be\label{bdpwr}
P(k)_{BD}=A_{s}\left(\frac{k}{k_{*}}\right)^{\left(n_{s}-1
\right)}\ee
where, $ A_{s} $, $ n_{s} $ are the scalar
amplitude, spectral index and $ k_{*} $ is the pivot scale.
It is straightforward to verify that this parameterization is quite generic to mimic almost all
the physical mechanisms and different models of inflation
(some of them have been discussed in the Introduction section (\ref{intro})), and it
has been widely used by the community to introduce the
generic (non-vacuum or non-standard) initial condition
into the primordial power spectrum\cite{Ross}. 
For detailed derivation of the power spectrum of this
form assumed for the BD/NBD case the interested reader can
go through the previous article by the same authors \cite{Chandra}.

Now, $\alpha_{k}$ and $\beta_{k}$ are known as
Bogolyubov coefficients (BCS), which parameterize
the initial state of the primordial fluctuations.
This form (\ref{nbdpwr}) of power spectrum arises,
when the general solution of the equation of motion,
usually known as Mukhanov-Sasaki equation, of the mode
function representing the scalar fluctuations, are
written in Hankel function basis, instead of considering
only the positive frequency solution, which is being
considered only for BD case. 
BCS $\alpha_{k}$ and $\beta_{k}$ satisfies a relation,
which follows directly from the Wronskian condition of
the mode function.

\be\label{norm}
\left |\alpha_{k}\right|^2 - \left |\beta_{k}\right|^2 = 1 
\ee 
One should note that the given normalization condition
is a generic one, which is satisfied by the BCS
irrespective of the physical processes are being addressed.
We will use the above normalization relation
(\ref{norm}) later on during further parameterization of the BCS as per the need of our analysis.

A brief discussion on BD vacuum and BCS is in order.
BD vacuum is defined as the lowest energy state of the
quantum fluctuations in de Sitter background, actually
it is an analogue of the Minkowski vacuum in flat spacetime,
as a result it has always been a favoured choice for
cosmological vacuum.
In Hankel function basis assuming $\alpha_{k} =1$ and
$\beta_{k}=0$ into the general solution of the mode
function, simply boils down to BD initial condition,
or simply in Hankel function basis BD vacuum is
parameterized by $\beta_{k}=0$.  Thus, a slightest
departure of the value of $\beta_{k}$ from $ 0 $ is
sufficient to conclude the existence of a non-trivial initial state.

As already pointed out, the true nature of the initial condition
of the inflaton fluctuations is still unknown to us.  In principle, the behaviour
of the initial state is dependent on scales, hence are BCS,
since the initial condition of the primordial
perturbations is characterized by BCS.
If the initial condition  is NBD in nature then only
the question of scale dependence is arising otherwise for BD
case the BCS are merely constants($\alpha_{k} =1$ and $\beta_{k}=0$). 
Theoretically the nature of the primordial initial state
or BCS has been investigated but depending upon different physical
scenarios the scale dependence and the amplitude of the BCS varies.
However, we still
do not even know what is the actual amplitude of the BCS,
so it is needless to say that, probing scale dependence
of the BCS in a model independent way is a far-fetched idea.
It may also be pointed out that,
although in principle BCS are scale dependant quantities that
may take different values for different
modes, BCS are itself small quantities and its variation
with scale would be even more smaller than that.  As a result,
we have safely assumed  the BCS as constants, without taking
into account their minor scale dependence, if any,  rather
considered their average values over scales as their amplitudes.
Further, in this article, we  refrain ourselves from considering any
particular form and amplitude of the BCS a-priori, rather we
plan to consider only the amplitude of the BCS for the time
being as the subject of focus in this present article, so
that we can  say something about the very nature of
the BCS directly from observations only, without considering
any specific model as such.

Satisfying the Wronskian condition (\ref{norm}), we further
parameterize  $\alpha_{k}$ and $\beta_{k}$ in a scale independent
form as argued above with keeping it in mind that BCS are
complex in nature:
 
 \[\alpha_k = \cosh \theta\]

 \[\beta_k= e^{-i \delta} \sinh \theta\]
 
\be \label{psnew}
P(k)_{NBD}  = A_{s}\left(\frac{k}{k_{*}}\right)^{\left(n_{s}-1\right)}
\left[1+ 2\sinh^2 \theta - 2 \sinh \theta \cosh \theta \cos\delta\right].
\ee

We will make use of this expression of NBD power spectrum
to model the dark matter power spectrum, which we will later
use to constrain the BCS from LSS observations in subsequent
sections of this article.

\section{From inflationary initial condition to structure formation}\label{sec3}
Thus far, we have discussed the parameterization of the
primordial power spectrum for non-trivial initial condition
in which the information about the initial state of
inflation is embedded in terms of BCS($\alpha_{k}$ and
$\beta_{k}$). As mentioned earlier the sole motive of
this analysis is to provide constraints on $\alpha_{k}$
and $\beta_{k}$ directly from LSS observations like SDSS
and Wiggle Z, for that purpose we need to relate the
primordial power spectrum with dark matter power spectrum.
For a consistent development of the analysis, let us do 
a very brief {\it textbook} review 
 of structure formation from inflationary
initial condition, that will be useful in the subsequent sections.
 For an extensive review one
can see \cite{Dodelson,Weinberg,Peebles}. The matter density fluctuations are related to primordial fluctuations through gravitational potential.  
The relation between primordial curvature
perturbations and primordial gravitational potential is as follows,
\be
\label{cppg}
\zeta(\mathbf{k}) =\frac{3}{2}\Phi_p(\mathbf{k}) 
\ee
where, $\Phi_p(\mathbf{k})$ is primordial gravitational
potential.
From perturbed Poisson and Euler equations for large scale and no radiation limit, one get the relation between primordial gravitational potential and the matter density perturbations,

\be
\label{Dmd}
\del(\mathbf{k},a)=\frac{3}{5} \frac{k^2T(k)D(a)}{\Omega_m H_0^2  }\Phi_p(\mathbf{k})
\ee

where, $ \Omega_m$ and $H_0 $ are respectively matter density
fraction and Hubble parameter at present time and $ \del(\mathbf{k},a) $ is the matter
density contrast. $ T(k) $ and $ D(a) $ are known as Transfer function and Growth factor respectively. Modifications in growth of different modes due to different
epoch of horizon re-entry is being described by Transfer
function ($ T(k) $). At sufficiently large scales transfer
function goes to one.
After the epoch of matter-radiation equality growth of the
density fluctuations are represented by Growth factor ($ D(a) $), which takes the value one at present epoch $(D(0)=1)$. 
 
Relating equation (\ref{nbdpwr}), (\ref{cppg})
and (\ref{Dmd})
we reach at: 
  
\be
\label{fnldmp}
 P_m(k,z)=\frac{8 \pi^2}{25}k \frac{T(k)^2 D(z)^2}
 {\Omega_m^2 H_0^4}
 P({k})_{NBD}
\ee

The process of the formation of dark matter fluctuations
from primordial perturbations is given by the above expression,
where $ P_m(k,z) $ stands for dark matter power spectrum.The transfer
function and the growth function can be evaluated numerically by CAMB
code \cite{CAMB} for $ \Lambda CDM $   Universe. 
The equation
(\ref{fnldmp}) represents the linear matter power spectrum. We
will use the power spectrum of the halo density field of the LSS
observations for our analysis to relate with SDSS and Wiggle Z observations. So, to construct the template of the halo density field
the prescription is very simple, we have to put the linear power spectrum
into halo fit following the instructions made by \cite{Smith,Takahashi}.
After that there has several real world effects like non-linearities,
redshift-space distortions to incorporate with the halo power spectrum.
Although the effect of redshift space distortions is not much effective
in linear region because, when the angle average of the power spectrum
is considered to address the mean effect of the redshift space
distortions it behaves as a mere amplitude  enhancement of the
power spectrum, which simply gets marginalized during the
marginalization of the unknown amplitude as nuisance parameter.
Now, there is another real-world effect, that is actually nonlinear
in nature, consequently it affects the scales of the power spectrum.
This nonlinear scale-dependent effect is widely known as Finger-of-God
effect. Although the Finger-of-God effect instigate the behaviour of
the scales, but the span of scale we are examining doesn't get strongly
influenced by this effect\cite{SDSS}, thereby we haven't include this
effect in our analysis.

Till date the observations draw inference on cosmological parameters by considering the initial condition as BD condition\cite{Planck}. Consequently, the transfer functions or the halofit models have
been developed thus far to match the present day LSS observations follows initial state as trivial BD vacuum. So, to carry the signature of a generic inflationary
initial state from primordial epoch to late time sky one may in principle have to slightly modify the transfer
function and the halo fit model, which will incorporate the behaviour of primordial initial state. Obviously
that won't be much different from the existing widely used transfer
functions and halo fit models as the correction due to non-standard initial vacuum would be small.

 In our preliminary analysis we have considered the usual transfer
function and halo fit model to evolve the primordial perturbations
to late time sky keeping in mind that the correction contribution due to non-trivial initial condition is small.

\section{Analysis and Data}\label{sec4} 
Let us now discuss 
the statistical methods used in the present analysis for finding
out possible  constraints on the  initial
condition. For our analysis, we have applied
Bayesian inference method with following four model
parameters:
 \[\left\lbrace \theta,~\delta,~\ln(10^{10}A_{s}),
 ~n_{s} \right\rbrace\]  
where, the model parameters are respectively,
parameterized BCS ($\theta,\delta$), the amplitude
of the primordial power spectrum ($ \ln(10^{10}A_{s}) $)
at the pivot scale $ k_{*}=0.05$/Mpc, and the spectral
index($ n_{s} $) of the primordial inflaton perturbations,
as demonstrated in Section \ref{sec2}.
For performing Bayesian inference analysis we have set
Gaussian prior with infinite variance for all the parameters.
Now to analyze BCS we will take shelter of MCMC simulation
technique, using the principle of Bayesian inference. For
that purpose we need to construct our likelihood
$ L(data;\alpha) $,
where $\alpha$ represents the model parameters
to be estimated (in the present analysis,
they are $ \theta,~\delta,~\ln(10^{10}A_{s})$ and
$ n_{s} $), and $ data $ stands for the observational data 
of our consideration (in the present problem,  WiggleZ and SDSS-DR7
LRG power spectrum data, taken separately). 
For our model the likelihood is defined as following:
\be
L(P^{obs},\alpha) 
=\frac{1}{(2\pi)^{n/2}|\textbf{Cov}|^{1/2}} /e^{-\chi^{2}/2}
\ee

where,
\be
\chi^{2} 
= (P^{obs}-P^{th})^{T}[\textbf{Cov}]^{-1}(P^{obs}-P^{th}),
\ee
Here $ P^{obs} $ is data vector for power spectrum
and $ P^{th} $ is our theoretical model template of the
power spectrum discussed in section \ref{sec3}, and
$ \textbf{Cov} $ stands for covariance matrix.

With the likelihood function and the data as defined above, we 
develop a code using the Markov Chain Monte Carlo (MCMC)
analysis based on Metropolis Hastings algorithm
in order to constrain our model parameters. To sample the
parameter space of the model parameters we use the random
walk metropolis algorithm and Cholesky decomposition.
We have monitored the convergence of the chains by
available MCMC convergence tests. Along with such
statistical convergence test, we have followed some visual
monitoring test too, in order to keep a track on the convergence
of the chain, using parameter trace plot, which gives
the plot of the value of the parameter versus iteration
for each parameter. To ensure the convergence and a
better sampling we ran all the chains starting from
random initial points. And we set the running time
accordingly to confirm the convergence of each chain
along with small relative errors of the means.

As mentioned earlier, the data we use are 
WiggleZ \cite{WZ} and SDSS data \cite{SDSS} from emission line galaxies
and luminous red galaxies respectively. The WZ galaxies
are basically the blue galaxies, known as star-forming
galaxies, usually avoid to stay around the centre of
cluster-size dark matter halos because it is being predicted
that these galaxies do not prefer to occupy the dense
regions of the halos\cite{Drinkwater} and the LRG are
massive galaxies that tend to populate the dense part of the
halos hence choose to remain at the centre of the
cluster-size dark matter halos\cite{Annis}. 
The power
spectrum constructed from the LRG catalog gives an
estimate of the power spectrum of the massive host halos
of the LRG. The two observations are quite different in
nature from various aspects such as their sensitivity
towards the nonlinearities and non-linear redshift space
distortions and most importantly the bias of the two
tracers are absolutely different. 
So,  we have performed two separate analysis by using these two datasets
separately, that will also help us in having a comparative analysis
of the results from two different LSS datasets.
The power spectrum we use  is the
power spectrum of the halo density field, which
are derived from the SDSS DR7 (LRG) survey\cite{SDSS}
and WiggleZ Dark Energy Survey\cite{WZ}. 
For our analysis
we have used the WiggleZ matter power spectrum data,
which involves four different redshift ranges respectively
$0.1 < z < 0.3$, $0.3 < z < 0.5$, $0.5 < z < 0.7$,
$0.7 < z < 0.9$, the effective redshift for these
four redshift bins are $ z=0.22,0.41,0.6,0.78$ while
the SDSS-DR7 LRG survey probes at redshift around
$ z\sim0.35 $ but being normalized to redshift $ z=0 $. 
For SDSS-DR7 survey the range of scale probed is from
0.02 $h$ $Mpc^{-1}$ to 0.2 $h$ $Mpc^{-1}$ \cite{SDSS}.
For WZ we have considered only the region within
$ k<k_{nl}=0.2$ $h$ $Mpc^{-1}$ of the data set,
the essence of which is that, it minimizes the
uncertainties cause due to non-linear
contributions, as has been explained in \cite{Riemer}.


\section{Results}

Having  demonstrated the methodology
and the data specifications of our analysis, we are now in a position 
to present our results obtained from Bayesian
inference technique using MCMC simulation. 
As has been explained, our
prime target in this article is to find the constraints
on the initial state of the primordial fluctuations in 
a model-independent way. In order to
do so we investigate the possible constraints on the
Bogolyubov coefficients, as BCS tell about the nature of
the initial states, and  individual initial states from
particular models can readily be cast in the form of BCS. We have
discussed, with elaborations, how to quantify initial state using
BCS in section \ref{sec2}. In a nutshell,
$ \alpha_{k}=1 $ and $ \beta_{k}=0 $ in equation
(\ref{nbdpwr}) or (\ref{norm}) represents BD initial
condition for inflationary quantum perturbations, any
value of $ \beta_{k}$ other than zero implies
deviation from BD initial condition. To satisfy the
Wronskian condition we further parameterize the $\alpha_k $
and $\beta_k $ as $\alpha_k = \cosh \theta$ and
$ \beta_k= e^{-i \delta} \sinh \theta $. Now, since
$ \beta_{k}=0 $ implies $\theta =0$, so any deviation
from $\theta =0$ is equivalent to deviation from
$\beta_{k}=0 $, which in turn predict the existence of
the NBD initial condition. Below we shall show the
results obtained from two LSS surveys separately.

\subsection{Constraints from Wiggle Z observations}
\begin{figure}[h] \label{fig1}
\center
\includegraphics[scale=0.70]{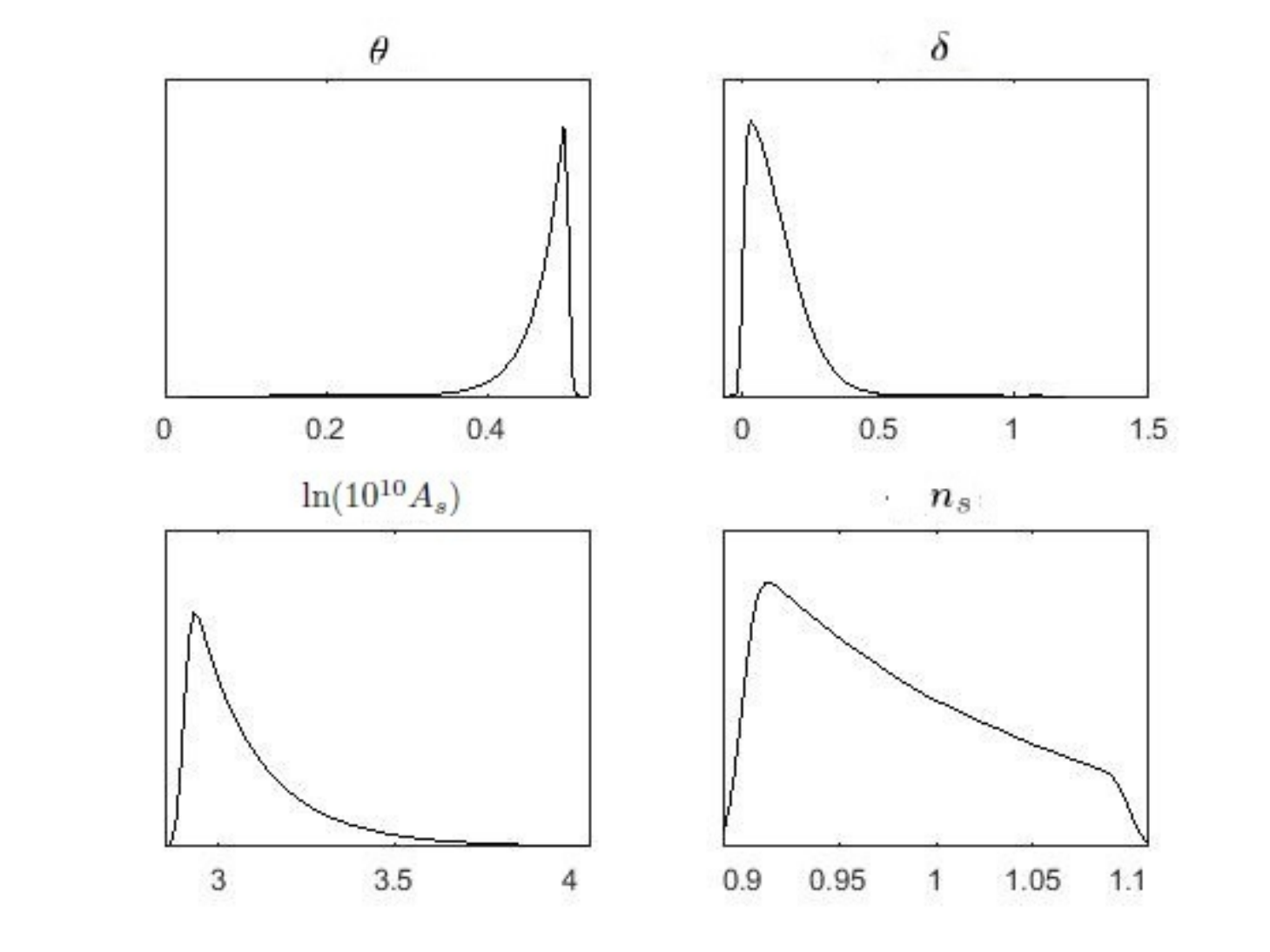}
\caption{The posterior of the model parameters
$\theta$, $\delta$, $\ln(10^{10}A_{s})$ and $ n_{s} $ are
being shown above for the Wiggle Z survey. }
\end{figure}

\begin{figure}[h]\label{fig2}
\center
\includegraphics[scale=0.70]{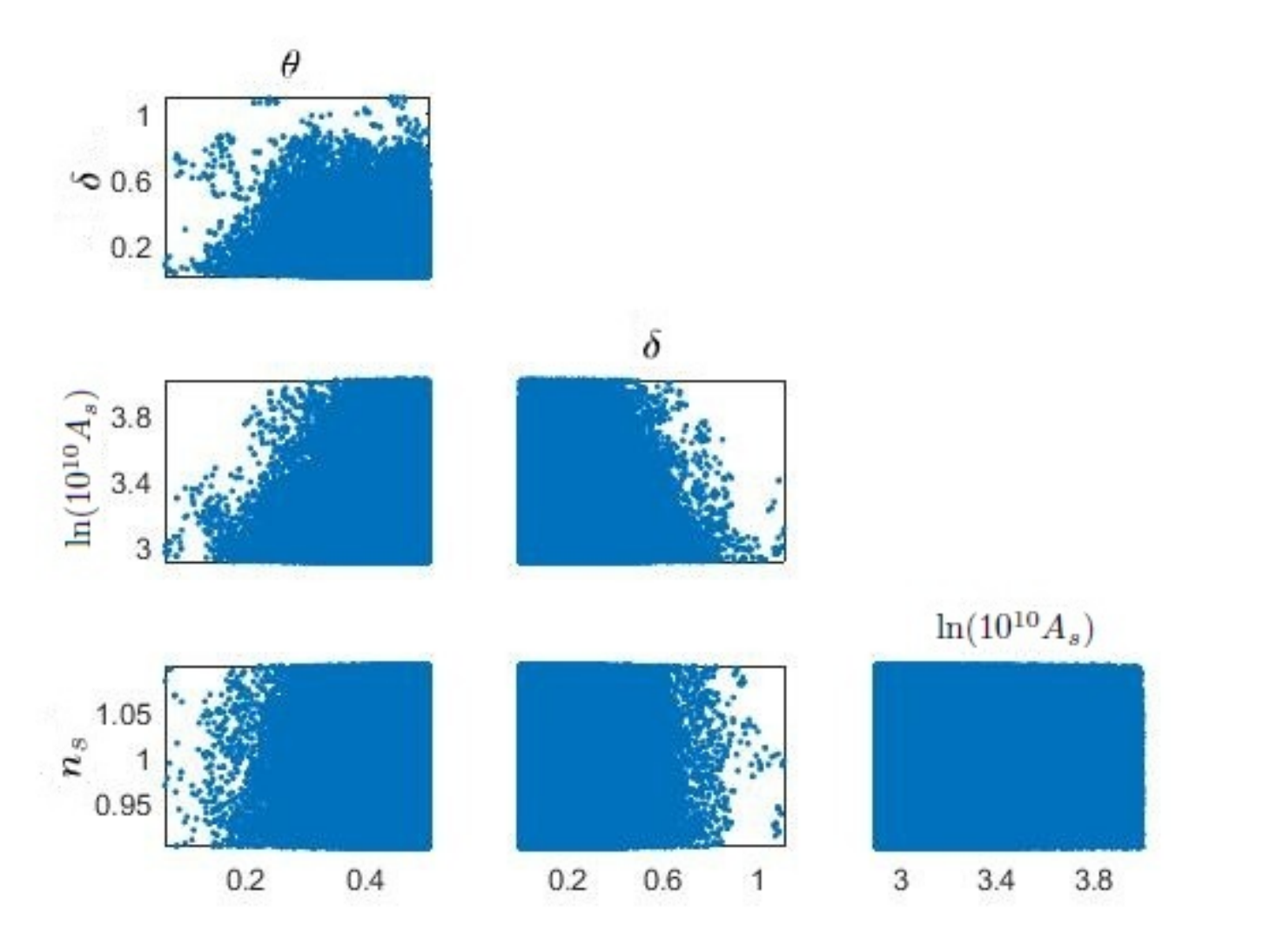}
\caption{This figure follows from the Wiggle Z survey, and
depicts the parameter space among four model
parameters $\theta$, $\delta$, $\ln(10^{10}A_{s})$ and $ n_{s} $.}
\end{figure}

Here we shall present the results obtained from the
Wiggle Z survey, by making use of the data discussed in details
in section \ref{sec4}.
The results appear in Figure 1(\ref{fig1}), Figure 2(\ref{fig2}) and
subsequently in Table 1(\ref{tab}). Figure 1(\ref{fig1}) and Figure 2(\ref{fig2})
respectively represent the posterior probability
and the scattered plot of the  parameters $\theta$,
$\delta$, $\ln(10^{10}A_{s})$ and $ n_{s} $ and in
Table 1(\ref{tab}) we have specified the mean values of the
model parameters along with their $1\sigma$ error bars
and corresponding $\chi^2$ for this data.
We will elaborate on the results in a bit details later in this section.

\subsection{Constraints from SDSS-DR7}
\begin{figure}[h] \label{fig3}
\center
\includegraphics[scale=0.70]{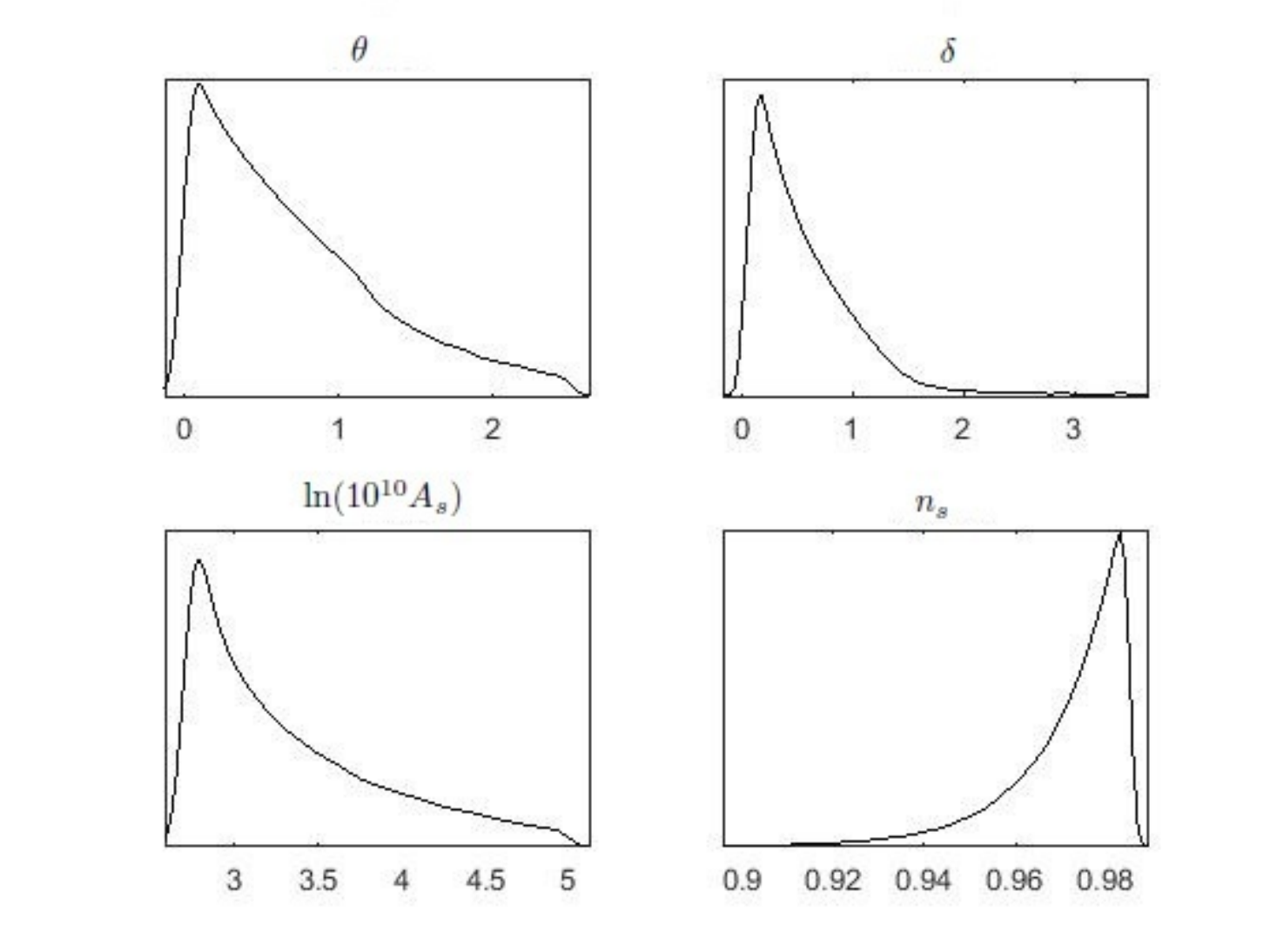}
\caption{The figure shows the posterior of $\theta$,
$\delta$, $\ln(10^{10}A_{s})$ and $ n_{s} $ obtained
from the SDSS-DR7 survey.  }
\end{figure}

\begin{figure}[h]\label{fig4}
\center
\includegraphics[scale=0.70]{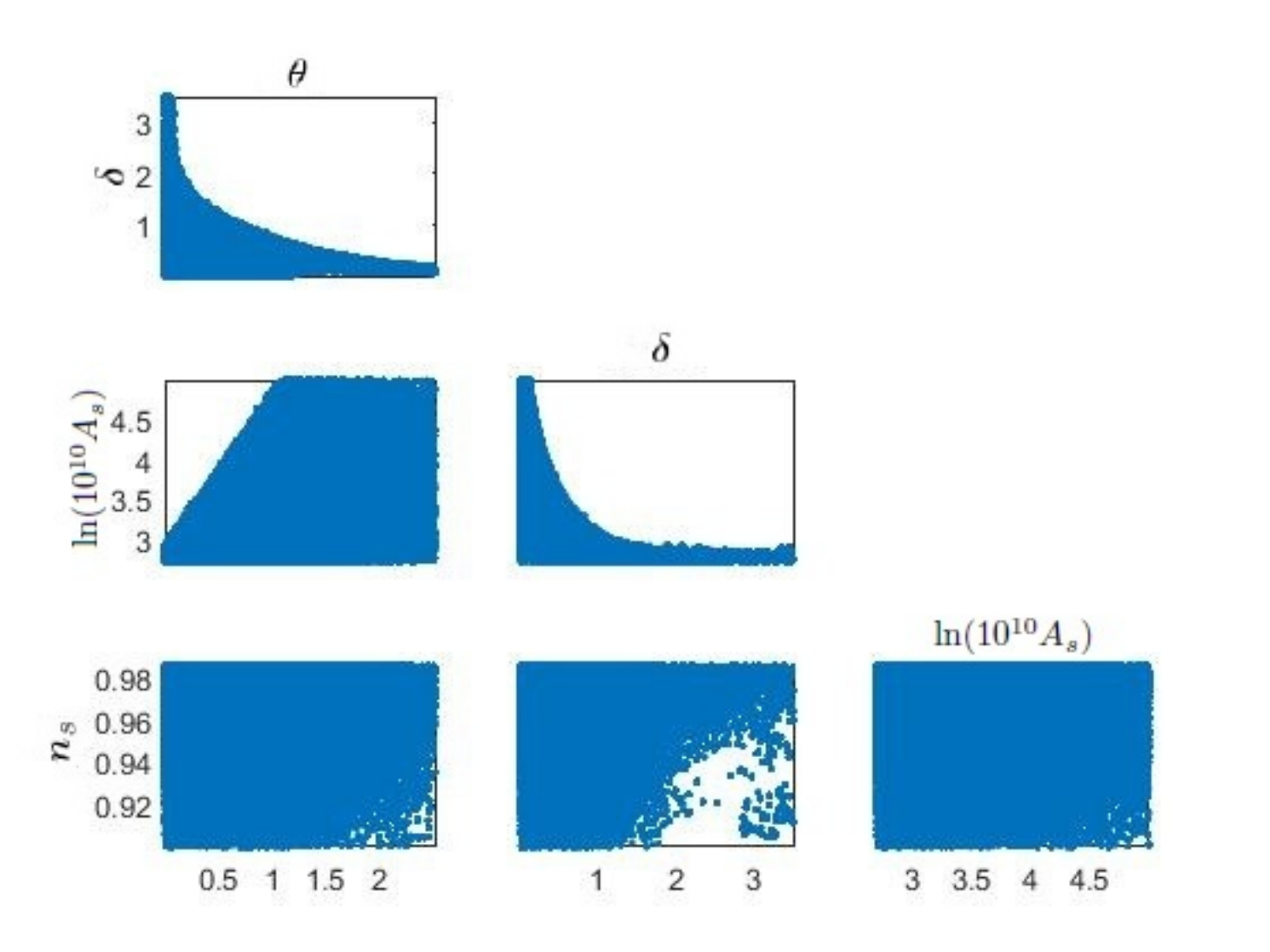}
\caption{This figure depicts the parameter space among
four model parameters $\theta$, $\delta$,
$\ln(10^{10}A_{s})$ and $ n_{s} $ for the SDSS-DR7 Survey.}
\end{figure}

Let us now demonstrate the results
obtained from the Sloan Digital Sky Survey - Data
Release 7 (SDSS-DR7) sample of Luminous Red Galaxies
(LRG)\cite{SDSS} as discussed in section \ref{sec4}. The results are shown in Figure 3(\ref{fig3}),
Figure 4(\ref{fig4}) and in the Table 1(\ref{tab}). As in the previous case,
Figure 3(\ref{fig3}) and
Figure 4(\ref{fig4}) respectively represent the posterior probability
and the scattered plot of the model parameters $\theta$,
$\delta$, $\ln(10^{10}A_{s})$ and $ n_{s} $ and in
Table 1(\ref{tab}) we have described the constraints on the
model parameters as the mean values with their
associated $1\sigma$ error bars, along with corresponding
$\chi^2$ for this data.

  \begin{table}
 \label{tab}
 \begin{tabular}{|l|l|l|l|l|l|l}
 \hline
 Data&$~~~~~~~~~~\theta$&$~~~~~~~~~\delta$&$~~~\ln(10^{10}A_{s})$&$~~~~~~~~~ n_{s} $ &$ ~~~~\chi^{2} $\\ \hline \hline
  SDSS-DR7 & $0.7217\pm 0.5879$ & $ 0.5684\pm 0.4942 $ & $3.3359\pm 0.5661$ & $  0.9711\pm 0.0136 $&$ 19.6387 $ \\ \hline
   Wiggle Z & $0.4682\pm 0.0322$ & $ 0.1226\pm 0.0962 $ & $3.0863\pm 0.1850$ & $  0.9771\pm 0.0548 $&$ 11.9005 $ \\ \hline
 \end{tabular}
 
 \caption{The above table designate the constraints on the values of the model parameters $\theta$,
$\delta$, $\ln(10^{10}A_{s})$ and $ n_{s} $ obtained
from SDSS-DR7 and Wiggle Z surveys. For all the parameters we report
    the mean values and the associated $1\sigma$ error bars.}
\end{table} 

A detailed analysis of the results obtained is in order.
First of all, the constraints on the primordial scalar
amplitude ($\ln(10^{10}A_{s})$) and the spectral
index($ n_{s}$) 
shown in Table 1(\ref{tab}) as obtained from these two individual
LSS datasets are consistent among themselves, and with the
latest CMB observations from Planck 2015 \cite{Planck}. The slight difference in
the shape of the posterior probabilities and the scattered
plots is due to the intrinsic differences in the two datasets, that might
have appeared either due to the differences in sensitivity towards nonlinearities
or due to the bias of the tracers.
Although both the observations are intrinsically very distinct from each other,
the striking resemblance between the results obtained from them is itself much
appealing.  Nevertheless, as is well known obtaining results from two different
observations also helps us in
narrowing down the parameter space of the model parameters.

As we can see from Table 1(\ref{tab}) and the corresponding 
plots, the results from both the surveys tell us
that the values of BCS ($\theta$, $\delta$) are non-zero 
at $1\sigma$ CL.
However, this does not essentially mean that BD vacuum
is ruled out at $1\sigma$, though it may wrongly point out to that.
At this stage of investigation, rather,
a more sensible statement is that there is a fair
chance for the initial state of the inflaton fluctuations
to be non-vacuum in nature.
There are two-fold reasons behind this statement :
First,  the BCS are scale-dependent, at least in principle,
so, without considering the actual scale dependence of the BCS,
if any,
 it would not be wise to comment conclusively
on whether or not  the BCS are NBD in nature  for all the modes.
It may so happen that
for some modes it is BD and for some particular modes
the initial state could be NBD. 
Whether or not BD is ruled out can conclusively be stated
only when one  
takes into account the scale-dependence, if any, of BCS,
which we did not do in our analysis. Consequently,
the results affect the amplitude of power spectrum only.
Therefore, any conclusive comment
on initial condition  before knowing its actual scale dependence
would be imprecise. However, a word of caution is that
whereas this can be done in principle, this may wash away 
our primary target, namely
 a model independent analysis of the whole scenario.

The second reason for refraining us
from making any conclusive comment is  due to the choice of the transfer
function and the halo fit model for the construction of
the theoretical template of the LSS power spectrum,
as discussed  in
section \ref{sec3}.
In order to
study the true imprints of the primordial initial state
on late time,  one should, in principle, propose an altogether new, reconstructed
(but presumably not too different from the standard one) 
transfer function from say, a modified halo fit model, which will
include the non-trivial effects of the primordial power spectrum for a
generic initial state.
But for our analysis we have assumed the standard transfer
function and halo fit model. Of course this is justified to a good extent
because of the small amplitude of the BCS, the correction
to the transfer function and the halo fit presumably would not be very much
significant. However, as long as one is not considering the reconstructed
transfer function and halo fit model, making any conclusive
comment would not be wise. So, all we can say at this stage is that
both BD and NBD initial states are allowed at $1\sigma$
 from Wiggle Z and SDSS-DR7 data for LSS.
This is in tune with our previous conclusions using CMB data
\cite{Chandra}.

\section{Summary and future directions}
In this article we make an attempt to investigate the
nature of the primordial initial state directly 
from observations. This is a part of the systematic,
stepwise development we have been planning to do, and, is in some sense a 
sequel of our previous article based on CMB data \cite{Chandra}.
This article in particular makes use of Large Scale Structure
data and attempts to find out possible constraints
on initial conditions therefrom, without giving priority to any specific
theoretical mechanism to set individual 
initial states. We build the generic
template of the primordial power spectrum by taking into account the
effect of both the non-standard initial state and the standard 
BD vacuum. Therefrom we develop the LSS power
spectrum using the standard transfer function for halo fit model. 
We then try to constrain this
generic power spectrum using two different LSS dataset, namely,
the WiggleZ Dark Energy Survey and the
Sloan Digital Sky Survey - Data Release 7 (SDSS-DR7) sample of
Luminous Red Galaxies (LRG), and  apply the Bayesian inference
algorithm to constrain the generic initial state. Our primary investigations
from
the statistical inference drawn results in a consistent power spectrum
and spectral index from the individual datasets which is in turn consistent
with latest CMB data from Planck 2015. Nevertheless, our analysis also gives us a hint that 
the initial state can well be a non-vacuum one due to 
the existence of  non-zero BCS from LSS data.

However, the template we construct 
in order to constrain the initial vacuum is limited in the sense
that we assume the BCS to be
constants for all the scales, which may not be absolutely true.
Such an assumption forbids us from making any strong
comment on the exact nature of the initial state, as
we can at best constrain the amplitude of the power spectrum using these
templates. We had to make such
a compromise in our analysis because in order to incorporate the scale dependence of BCS, 
one may have to 
take shelter of  particular physical process depending on models.
Whereas this can be done in principle, this may in turn wash away the basic
essence of the primary target
of the present paper, which  was to
do a model independent analysis of the scenario.
However, we are already exploring situations where one can 
build a somewhat phenomenological
scale dependent structure of the initial state, and make a 
more or less generic comment therefrom.
This is  our next venture in this direction
of probing the inflationary initial state. 
Nevertheless, in order to extend our analysis towards a 
systematic,
stepwise development of the problem, one needs to constrain the initial state
using other dataset as well. One can also forecast
on different observations based on a similar analysis.
Further, it was pointed out in the article that, for simplicity,
we have assumed the standard transfer function with halo fit model
in our analysis. To be honest, in this scenario one needs to do a 
reconstruction of transfer function to reflect any non-trivial
effect of generic initial condition on the primordial power spectrum.
We plan to explore in these directions in near future.

To conclude, the article is indeed  a small  step forward
towards answering one of the yet-unanswered fundamental questions
related to inflationary paradigm, namely the nature of inflationary initial state,
using LSS data. However, much work needs to be done  before
one can make 
any decisive comment on the exact nature of the primordial initial state.


\section*{Acknowledgments}

DC thanks ISI Kolkata for financial support through
Senior Research Fellowship. We gratefully
acknowledge the computational facilities of
ISI Kolkata.



\begin{thebibliography}{}

\bibitem{Wmap} C. L. Bennett et al. [WMAP collaboration],  [arXiv:1212.5225] [astro-ph.CO].

\bibitem{Planck}
  P. A. R. Ade et al. [Planck collaboration],
   [arXiv:1502.01589] [astro-ph.CO].\\





\bibitem{Guth}
A. H. Guth, S. Y. Pi, Phys.Rev.Lett. 49, 1110 (1982);\\
A. A. Starobinsky, Phys.Lett.B117, 175 (1982);\\
S. W. Hawking, Phys.Lett.B115, 295 (1982);\\
J. M. Bardeen, P. J. Steinhardt, and M. S. Turner,
 Phys.Rev.D28, 679 (1983);\\
V. F. Mukhanov and G. V. Chibisov, JETP Lett.
 33, 532 (1981).\\ 

\bibitem{Mirbabayi}
M. Mirbabayi, L. Senatore, E. Silverstein, and
 M. Zaldarriaga, Phys.Rev.D 91 (2015)
  063518 [arXiv:1412.0665] [hep-th].\\

\bibitem{Bartolo}
N. Bartolo, E. Komatsu, S. Matarrese, and 
A. Riotto, Phys.Rept. 402 (2004) 
103 [arXiv:astro-ph/0406398].\\

\bibitem{Martin}
J. Martin and R. H. Brandenberger, 
Phys.Rev.D 63 (2001) 123501 [arXiv:hep-th/0005209].\\

\bibitem{Chandra}
D. Chandra and S. Pal, Class.Quant.Grav.
 35, 015008 (2017)  [arXiv:1606.09098] [hep-th].\\

\bibitem{Davies}
 T. S. Bunch and P. C. W. Davies, Proc. Roy. Soc. Lond. A
360, 117 (1978).\\

\bibitem{Easther}
R. Easther, B. R. Greene, W. H. Kinney, 
G. Shiu, Phys.Rev.D64, 103502 
(2001) [arXiv:hep-th/0104102].\\


\bibitem{Wang} X. Chen, Y. Wang, JCAP 1407, 004 (2014)
  [arXiv:1306.0609] [hep-th].\\


\bibitem{chen2}
 X. Chen, JCAP 1012:003.2010 [arXiv:1008.2485] [hep-th].\\


 \bibitem{Gasperini}
  M. Gasperini, M. Giovannini and G. Veneziano, 
  Phys.Rev.D48:439-443,1993 [arXiv:gr-qc/9306015];\\
 K. Bhattacharya, S. Mohanty and R. Rangarajan,  
  Phys.Rev.Lett.96:121302,2006 [arXiv:hep-ph/0508070];\\
 K. Bhattacharya, S. Mohanty and A. Nautiyal, 
  Phys.Rev.Lett.97:251301,2006 [arXiv:astro-ph/0607049];\\
  I. Agullo and L. Parker,  
  Phys.Rev.D83:063526,2011 [arXiv:1010.5766] [astro-ph.CO];\\
  M.~Giovannini, 
  Phys.Rev.D88, no. 2, 021301 (2013) [arXiv:1304.4832] [astro-ph.CO];\\
 
 \bibitem{danielsson}
U. H. Danielsson, Phys.Rev.D66 (2002) 023511
 [arXiv:hep-th/0203198];\\
J. Martin and R. H. Brandenberger, Phys.Rev.D63:123501,
 2001 [arXiv:hep-th/0005209];\\
A. Kempf, Phys.Rev.D63 (2001) 083514 [arXiv:astro-ph/0009209];\\

  Phys.Rev.D67 (2003) 063508 [arXiv:hep-th/0110226];
   Phys.Rev.D66:023518.2002 [arXiv:hep-th/0204129];\\
N. Kaloper, M. Kleban, A. E. Lawrence, S. Shenker,
 Phys.Rev.D66  (2002) 123510 [arXiv:hep-th/0201158];\\
K. Schalm, G. Shiu, J. P. V. D. Schaar,
 JHEP 0404:076,2004 [arXiv:hep-th/0401164]; \\
 261-273 [arXiv:gr-qc/0411056];\\
A. Ashoorioon, A. Kempf, R. B. Mann, Phys.Rev. D71
(2005) 023503 [arXiv:astro-ph/0410139];\\
A. Ashoorioon, J. L. Hovdebo, R. B. Mann, Nucl.Phys.
 B727 (2005) 63-76 [arXiv:gr-qc/0504135];\\
L. Hui and W. H. Kinney, Phys.Rev.D65 (2002) 103507
 [arXiv:astro-ph/0109107];\\
A. Ashoorioon, K. Dimopoulos, M. M. Sheikh-Jabbari
 and G. Shiu, JCAP 02 (2014) 025 [arXiv:1306.4914];\\
A. Ashoorioon, K. Dimopoulos, M. M. Sheikh-Jabbari
 and Gary Shiu, Physics Letters B 737 (2014) 98.\\


\bibitem{shiu} 
  G. Shiu and J. Xu, Phys.Rev.D84,103509 (2011)
   [arXiv:1108.0981];\\
  A. Ashoorioon, A. Krause, K. Turzynski,
   JCAP 0902:014,2009 [arXiv:0810.4660] [hep-th];\\
  A. Ashoorioon, A. Krause, [arXiv:hep-th/0607001].\\

\bibitem{Alvarez} M. Alvarez et al.,
  [arXiv:1412.4671] [astro-ph.CO].\\



 \bibitem{Carrasco} 
  J. J. M. Carrasco, M. P. Hertzberg and
   L. Senatore, JHEP 1209 (2012) 082,
    [arXiv: 1206.2926] [astro-ph.CO].\\

\bibitem{SDSS} A. B. Reid et. al., 
Mon. Not. R. Astron. Soc 404 (2010) 
60-85 [arXiv:0907.1659] [astro-ph.CO].\\

\bibitem{WZ} D. Parkinson et. al., 
Phys. Rev. D. 86 (2012) 103518 [arXiv:1210.2130] [astro-ph.CO].\\



  

 
 










\bibitem{Weinberg} S. Weinberg, Cosmology, Oxford Univ. Press (2008).\\



\bibitem{Peebles} P. J. E. Peebles,  1980, 
Large-Scale Structure of the Universe 
(Princeton: Princeton Univ. Press).\\

\bibitem{Dodelson} S. Dodelson, 2003, 
Modern Cosmology (New York: Academic press).\\




\bibitem{Smith}
R. E. Smith et al. Mon.Not.Roy.Astron.Soc. 
341 (2003) 1311 [arXiv:astro-ph/0207664].\\

\bibitem{Takahashi}
R. Takahashi et al. [arXiv:1208.2701] [astro-ph.CO].\\



 
\bibitem{CAMB} A. Lewis, A. Challinor and A. Lasenby,
 Astrophys. J. 538, 473 (2000)  [arXiv:astro-ph/9911177].\\
 

 
  



 
  
 
 
  
  
  \bibitem{Ross} 
 R. O'Connell and R. Holman, [arXiv: 1109.1562] [hep-th].\\
 
 \bibitem{Drinkwater} 
 M. J. Drinkwater et al., MNRAS 401, 1429 (2010),
  [arXiv: 0911.4246] [astro-ph.CO].\\

\bibitem{Annis} 
D. J. Eisenstein et al., AJ 122, 2267 (2001),
  [arXiv: astro-ph/0108153].\\
  
\bibitem{Riemer} 
 S. Riemer-Sorensen et al., Phys. Rev. D85,
  081101 (2012), [arXiv: 1112.4940] [astro-ph.CO].\\
  













\end{thebibliography}
\end{document}